\documentclass[twocolumn,tighten,times]{aastex62}

\usepackage{graphics,epsf}
\usepackage{amsmath}                
\usepackage{amsfonts}               
\usepackage{amssymb}                
\usepackage{epsfig}
 \usepackage{epstopdf}
\usepackage{multirow}
\usepackage[para,online,flushleft]{threeparttable}
\def \kms{~\rm{km~s^{-1}}}

\def \msyr{~\rm{M_{\odot}}~\rm{yr^{-1}}}

\def \s{~\rm{s}}
\def \km{~\rm{km}}

\def \K{~\rm{K}}

\def \au{~\rm{au}}
\def \erg{~\rm{erg}}

\def \yr{~\rm{yr}}

\def \etc{$\eta$~Car}

\def \rmModot{~\rm{M_{\sun}}}
\def \rmRodot{~\rm{R_{\sun}}}
\def \rmLodot{~\rm{L_{\sun}}}

\usepackage{xcolor}
\definecolor{redak}{rgb}{0.9,0.15,0.05}

\begin{document}

\title{\Large{The orientation of Eta Carinae and the powering mechanism of intermediate luminosity optical transients (ILOTs)}}

\author[0000-0002-7840-0181]{Amit Kashi}
\affil{Department of Physics, Ariel University, Ariel, POB 3, 40700, Israel}
\email{kashi@ariel.ac.il}
\author[0000-0003-0375-8987]{Noam Soker}
\affil{Department of Physics, Technion -- Israel Institute of Technology, Haifa 32000, Israel}
\affil{Guangdong Technion Israel Institute of Technology, Shantou, Guangdong Province, China}
\email{soker@physics.technion.ac.il}

\shorttitle{Orientation of $\eta$ Car and powering ILOTs} 
\shortauthors{A. Kashi \& N. Soker}

\begin{abstract}

Contrary to recent claims, we argue that the orientation of the massive binary system Eta Carinae is such that the secondary star is closer to us at periastron passage, and it is on the far side during most of the time of the eccentric orbit. The binary orientation we dispute is based on problematic interpretations of recent observations. Among these observations are the radial velocity of the absorption component of \ion{He}{1} P-Cyg lines, of the \ion{He}{2}~$\lambda4686$ emission line, and of the Br~$\gamma$ line emitted by clumps close to the binary system. We also base our orientation on observations of asymmetric molecular clumps that were recently observed by ALMA around the binary system, and were claimed to compose a torus with a missing segment. The orientation has implications for the modeling of the binary interaction during the nineteenth century Great Eruption (GE) of Eta Carinae that occurred close to periastron passage. The orientation where the secondary is closer to us at periastron leads us to suggest that the mass-missing side of the molecular clumps is a result of accretion onto the secondary star during the periastron passage when the clumps were ejected, probably during the GE. The secondary star accreted a few solar masses during the GE and the energy from the accretion process consists the majority of the GE energy. This in turn strengthens the more general model according to which many intermediate-luminosity optical transients (ILOTs) are powered by accretion onto a secondary star.
\end{abstract}


\section{INTROCUTION}
\label{sec:intro}

In the controversy on the orientation of the Eta Carinae (\etc) binary system there are, literally, 180 degree opposite views.
The majority holds that, during periastron passages the primary star is
closer to us ($\omega \simeq 240^\circ$--$270^\circ$; e.g.,
\citealt{Nielsenetal2007, Daminelietal2008b, Parkinetal2009,
Maduraetal2012, Richardsonetal2015, Richardsonetal2016, Teodoroetal2016} and many others), while the other side holds the view that during periastron passage the secondary star is at its closest location to us ($\omega \simeq
90^\circ$; e.g., \citealt{Abrahametal2005, Falcetaetal2005,
AbrahamFalceta2007, KashiSoker2008, KashiSoker2009b, KashiSoker2009d, KashiSoker2011, KashiSoker2016Massive, Kashietal2011, SokerKashi2012, Tsebrenkoetal2013}).
   
The interpretation we prefer with $\omega \simeq 90^\circ$, comes from detailed analysis and modeling of a plethora of observations, including many absorption and emission lines from many parts of the binary system (primary, secondary, their winds, the colliding winds structure, and gas located further out of the binary orbit; \citealt{KashiSoker2009b,KashiSoker2009d}), observations of the hydrogen column density toward the binary system as deduced from X-ray absorption \citep{KashiSoker2009b,KashiSoker2009d}, and emission from blobs in the vicinity of the binary system such as the Weigelt blobs environment \citep{SokerKashi2012}.
We discussed in length in our previous papers why we believe the conclusions in the works listed above that suggest an orientation of $\omega \simeq 240^\circ$--$270^\circ$ have severe flaws.

As more observations and publications suggesting the opposite orientation to our view have been published lately (references below), we address the orientation question again.
We do so by examining each and every observation and explaining why using it to deduce the opposite orientation encounters difficulties. 

The question of the orientation of $\eta$ Car has a wider scope 
than merely the accidental position of the binary system on the sky.
As the same orientation (almost) of the binary system was involved in the shaping and determination of the circumbinary gas distribution, locations of clumps, filaments, and voids that were ejected in the Great Eruption (GE) of 1837--1856 (e.g., \citealt{DavidsonHumphreys2012} and references therein), 
it has further implications on the study of how binary systems power intermediate-luminosity optical transients (ILOTs). These are various kinds of eruptive stellar events with peak luminosity below those of
supernovae and above those of novae, and share properties with LBV giant eruption \cite{KashiSoker2016}. 

The paper is organized such that each section confronts an interpretation of different observations (all of which, according to their authors, lead to the orientation we oppose).
In section \ref{sec:lines} we disprove interpretation of observations of spectral lines.
In section \ref{sec:fan} we show that the observed fan in line-of-sight velocity slices of the Br~$\gamma$ line does not indicate anything about the orientation.
In section \ref{sec:torus} we present a different conclusion regarding the recently studied CO torus around $\eta$ Car with the Atacama Large Millimeter/submillimeter Array (ALMA).
We summarize in section \ref{sec:summary} where we also discuss the implications of our deduced orientation on the powering mechanism of ILOTs.

\section{SPECTRAL LINES}
\label{sec:lines}
\subsection{Absorption lines}
\label{subsec:abslines}

Radial velocity measurements of spectral lines have been used to try and deduce the orbital orientation of $\eta$ Car (e.g., \citealt{Nielsenetal2007, Daminelietal2008b, Mehneretal2010, Mehneretal2011a, Richardsonetal2015, Richardsonetal2016}). In our most recent paper on the subject \citep{KashiSoker2016Massive} we have already dealt with all observations and results published before 2016. In that paper we analyzed in depth the results of \cite{Richardsonetal2015}, and presented our arguments for the orientation of the secondary being on the far side from us at apastron ($\omega=90^\circ$). In particular, we showed that, by taking $\omega=90^\circ$ and assuming that the velocity of the trough in the P~Cyg profiles come from the secondary star, we can much better fit the velocity variations along the orbital period than with the model of \cite{Richardsonetal2015}, which assumes $\omega \simeq 240^\circ$--$270^\circ$. 
 
In a more recent paper \cite{Richardsonetal2016} present more observations from which they deduce velocities of absorbing segments before, during, and after the 2014.6 periastron passage.  They present the velocity variation of the \ion{He}{1}~$\lambda5876$ line, and fit a velocity curve that is appropriate to the velocity of the primary star in the orientation of the secondary star, being closer to us near apastron $\omega \simeq 243^\circ$ (middle panel in their figure 4). However, it is not clear that this fit is as good as might seem at first sight.  
They find for the 2014.6 event that the velocity variation between the epochs before and after the event is only about $\Delta v= 180  \km \s^{-1}$, compared with $\Delta v= 380 \km \s^{-1}$ that \cite{Nielsenetal2007} find for the 1998.0 and 2004.3 events. It is clear that the fitting used by \cite{Richardsonetal2016} cannot fit either the observations of \cite{Nielsenetal2007} or those of \cite{Richardsonetal2015}, as we showed in \citep{KashiSoker2016Massive}. 

Our explanation for the above disagreement on whether the fitting of the $\omega \simeq 243^\circ$ can work or not is twofold. (1) We note that \cite{Richardsonetal2016} do not have observations in the phase range $0.013$--$0.036$. Their points at phase 0.01 and at phase 0.035 have higher velocities (more red shifted) than their fit. From the observations of the same spectral line by \cite{Nielsenetal2007} the velocities in the phase range $0-0.035$ are $100 \km \s^{-1}$ more red shifted than the velocity of \cite{Richardsonetal2016} at phase 0.035. This holds for other \ion{He}{1} lines observed by \cite{Nielsenetal2007} and by \cite{Richardsonetal2015}. 
(2) Just before periastron, \cite{Richardsonetal2016} present velocities with an amplitude smaller than what \cite{Nielsenetal2007} and \cite{Richardsonetal2015} find. The last two papers present velocities that are blue shifted by about $60 \km \s^{-1}$ more than what \cite{Richardsonetal2016} present in the middle panel of their figure 4. However, by examining the right panel of their figure 6, we see that the blueshift absorption indeed reaches indeed higher values, more than $100 \km \s^{-1}$ than presented in their figure 4. 
\cite{Richardsonetal2016} simply identify this high velocity as a different component from that presented in their figure 4. We find no justification for that. The line profile could be well fitted by one absorbing component and few emission components
\citep{Nielsenetal2007}. 

We conclude that \cite{Richardsonetal2016} were able to fit the velocity variation of the \ion{He}{1}~$\lambda5876$ P Cygni line with their preferred orientation of $\omega \simeq 243^\circ$ because they were missing observations in the phase range $0.01-0.035$ and because they ignored what they term ``component 4''. When these are included, the fitting with the orientation where the secondary star is closer to us at apastron $\omega \simeq 270^\circ$ is very bad, as we have shown before \citep{KashiSoker2016Massive}. 
For the other properties of the velocity, \cite{Richardsonetal2016} present only qualitative explanation. As they completely ignore our arguments presented in the early paper \citep{KashiSoker2016Massive} and did not refute any of our claims, there is no basis for further discussion here, and our previous arguments from \citep{KashiSoker2016Massive} remain valid.

\subsection{The \ion{He}{2}~$\lambda4686$ line}
\label{subsec:HeII4686line}

Though many of the observations of $\eta$ Car have received different interpretations, there is consensus that the \ion{He}{2}~$\lambda4686$ line is the least understood observation of the system. 
The \ion{He}{2}~$\lambda4686$ line has special significance because it responds to very soft X-rays and the ionizing UV radiation field of both stars \citep{Davidsonetal2015}.
The line's equivalent width (EW) has three characteristic peaks across periastron, which not only received different interpretations but also were not measured with the same intensity by different telescopes, what raised arguments regarding the validity of some observations \citep{Davidsonetal2015}.
The line also shows secular changes from one spectroscopic event to the next, and serves as an indicator of variations in the intensity of the primary wind.
Since the primary photosphere is located in the wind, decreasing wind densities imply a smaller photospheric radius, higher radiation temperatures, and higher ionization states \citep{Davidsonetal2015}.

\cite{Mehneretal2011b} have observed the  \ion{He}{2}~$\lambda4686$ emission 
both directly and reflected from the poles from the location known as FOS4.
They found both the EW and the radial velocities to be very similar when viewed from the two different directions.
They concluded that the velocities of the line are not simply related to the orbital motion of the secondary star.

\cite{Teodoroetal2012} interpreted the 2009 X-ray light curve of $\eta$ Car and the \ion{He}{2}~$\lambda4686$ EW curve as a result of clumping of the primary wind, which causes flare-like behavior of both light curves.
They assumed that the primary source of the emission of this line is located at the apex of the wind-wind collision, and that the X-ray emission ionizes the He$^{+}$ ions. 
The first authors to suggest that the \ion{He}{2}~$\lambda4686$ emission comes from the winds collision region were \cite{Martinetal2006}. This was also the first paper to provide an analysis of the special 
radiative-transfer physics of \ion{He}{2} in $\eta$ Car.
An earlier paper by \cite{SteinerDamineli2004} claimed the \ion{He}{2}~$\lambda4686$ emission comes from an
inner part of the secondary wind, allegedly ionized by radiation from the shock fronts.
However, the estimate given there for the luminosity in the line was by 2--4 orders of magnitude too large.

\cite{Teodoroetal2012} attributed the late maximum that occurs when the X-ray has already gone to minimum (also known as peak 3 or P3) to a `collapse' of the colliding winds region onto regions of the secondary stellar wind, where it is still being accelerated, and later onto the surface of the secondary star, as a result of radiative inhibition
of the acceleration of the secondary wind to its terminal velocity.
This idea is similar to the models later described in \cite{Parkinetal2009} and \cite{Corcoranetal2010}.
As \cite{Teodoroetal2012} noted, accretion of clumps was suggested by \cite{Akashietal2006}, and was used to model the \ion{He}{2}~$\lambda4686$ line in \cite{SokerBehar2006} and \cite{Soker2007}.
However, in contrast, \cite{SokerBehar2006} postulated that this line originates in the acceleration zone of the secondary wind rather than the colliding winds region.
\cite{Teodoroetal2012} claimed that accretion is not a necessary condition to describe the observed phenomena in the \ion{He}{2}~$\lambda4686$ line, and the `collapse' is sufficient.
In either case, the question regarding the occurrence of accretion close to periastron passage is no longer current, as simulations by \cite{Akashietal2013} and \cite{Kashi2017} unambiguously showed.
It was also shown that accretion cannot be prevented by radiative braking, as claimed by \cite{Maduraetal2013}.

\cite{Teodoroetal2016} also studied the \ion{He}{2}~$\lambda4686$ line, this time for the 2014.6 periastron passage, and again suggested the same model as in \cite{Teodoroetal2012}.
The model that \cite{Teodoroetal2016} suggest for the \ion{He}{2}~$\lambda4686$ line fails on two counts. First is the argument of the photon number that \cite{SokerBehar2006} presented ten years earlier and which was completely ignored by \cite{Teodoroetal2016}. Let us briefly repeat here this argument. 
The observed X-ray luminosity at peak is $\simeq 2.5 \times 10^{35} \erg \s^{-1}$ (e.g., \citealt{Corcoranetal2017}), but the intrinsic X-ray emission can be as high as $\approx 4 \times 10^{36} \erg \s^{-1}$ \citep{SokerBehar2006}. For a plasma at a temperature of $\approx 5 \times 10^6 \K$, the number rate of photons that might ionize the He$^{+}$ ions is $\dot N_{\rm X} \approx 10^{46} \s^{-1}$. 
The \ion{He}{2}~$\lambda4686$ line at maximum has a luminosity of $L_{\rm HeII,max} \approx 300 \rmLodot$ (e.g., \citealt{Teodoroetal2016}), which amounts to a photon number per second of $\dot N_{\rm HeII} \approx 3 \times 10^{47} \s^{-1} \approx 30 \dot N_{\rm X}$. Since not all the X-ray photons will be absorbed by He$^{+}$ ions, we conclude that the ionization rate by the X-ray photons of the colliding winds is short by about two orders of magnitude from what is required for the formation of the \ion{He}{2}~$\lambda4686$ line.
Even if \citep{SokerBehar2006} underestimated the ionizing flux by a factor of a few, it is still an order of magnitude from that required.
 
The second problem of the model that \cite{Teodoroetal2016} propose is that just before periastron passage the \ion{He}{2}~$\lambda4686$ line is highly blue-shifted. This implies that just before periastron passage the material that emits the line is moving toward us at higher velocity than during the other orbital phases. But according to their orientation of $\omega \simeq 234^\circ-252^\circ$, the apex of the winds collision is moving away from us as the system approaches periastron. 
 
Admittedly, the explanation for the origin of the \ion{He}{2}~$\lambda4686$ emission line is complicated. The suggestion made by \cite{SokerBehar2006} that the line originates in the acceleration zone of the secondary wind, stands up better than that of \cite{Teodoroetal2016}, but it is also not free from problems. It is not clear that the  origin in the zone of the secondary wind can explain the velocity variation along the orbit.
Here we limit ourselves to firmly conclude that the arguments presented by \cite{Teodoroetal2016} to support an orbital orientation of $\omega \simeq 234^\circ-252^\circ$, i.e., the secondary star toward us at apastron, do not hold. 

\section{THE FAN-SHAPED STRUCTURE}
\label{sec:fan}

\cite{Weigeltetal2016} observed $\eta$ Car in the Br~$\gamma$ line and presented their observations at different line-of-sight velocities. This binning of the line-of-sight velocity reveals a structure in the velocity range of $-140 \kms$ to $-376 \kms$ which \cite{Weigeltetal2016} identify as a fan-shaped structure.
They argued that the south-east direction of the opening of the fan matches the expectation according to the simulations of the colliding winds made by \cite{Maduraetal2013}. 
This led \cite{Weigeltetal2016} to conclude that their observations support an orientation where the secondary star is away from us at periastron ($\omega \simeq 240-270^\circ$).  
 
\begin{figure}
\centering
\includegraphics[trim= 0.1cm 0.0cm 0.2cm 0.1cm,clip=true,width=0.82\columnwidth]{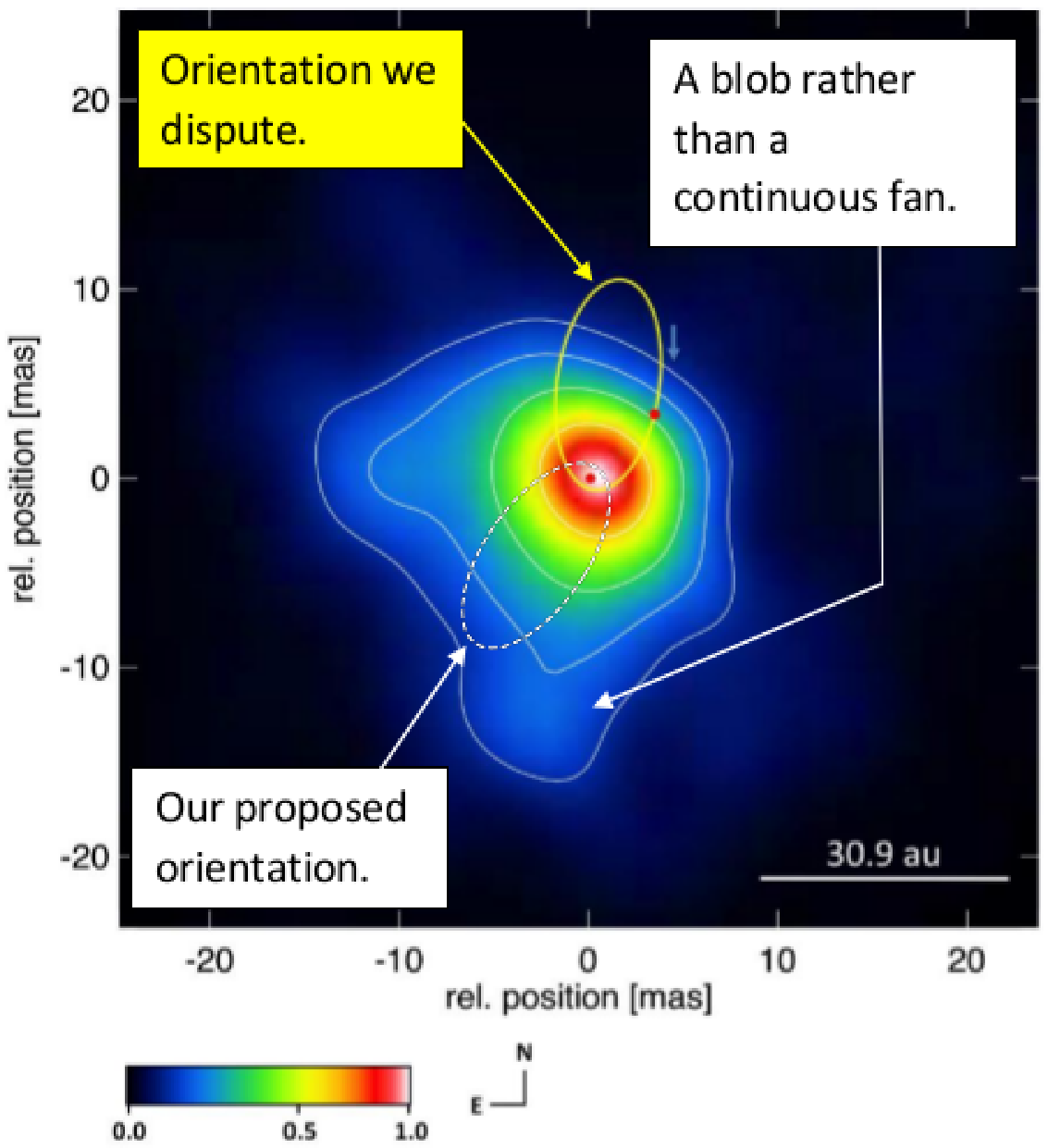}  
\includegraphics[trim= 0.0cm 0.0cm 0.0cm 0.0cm,clip=true,width=0.82\columnwidth]{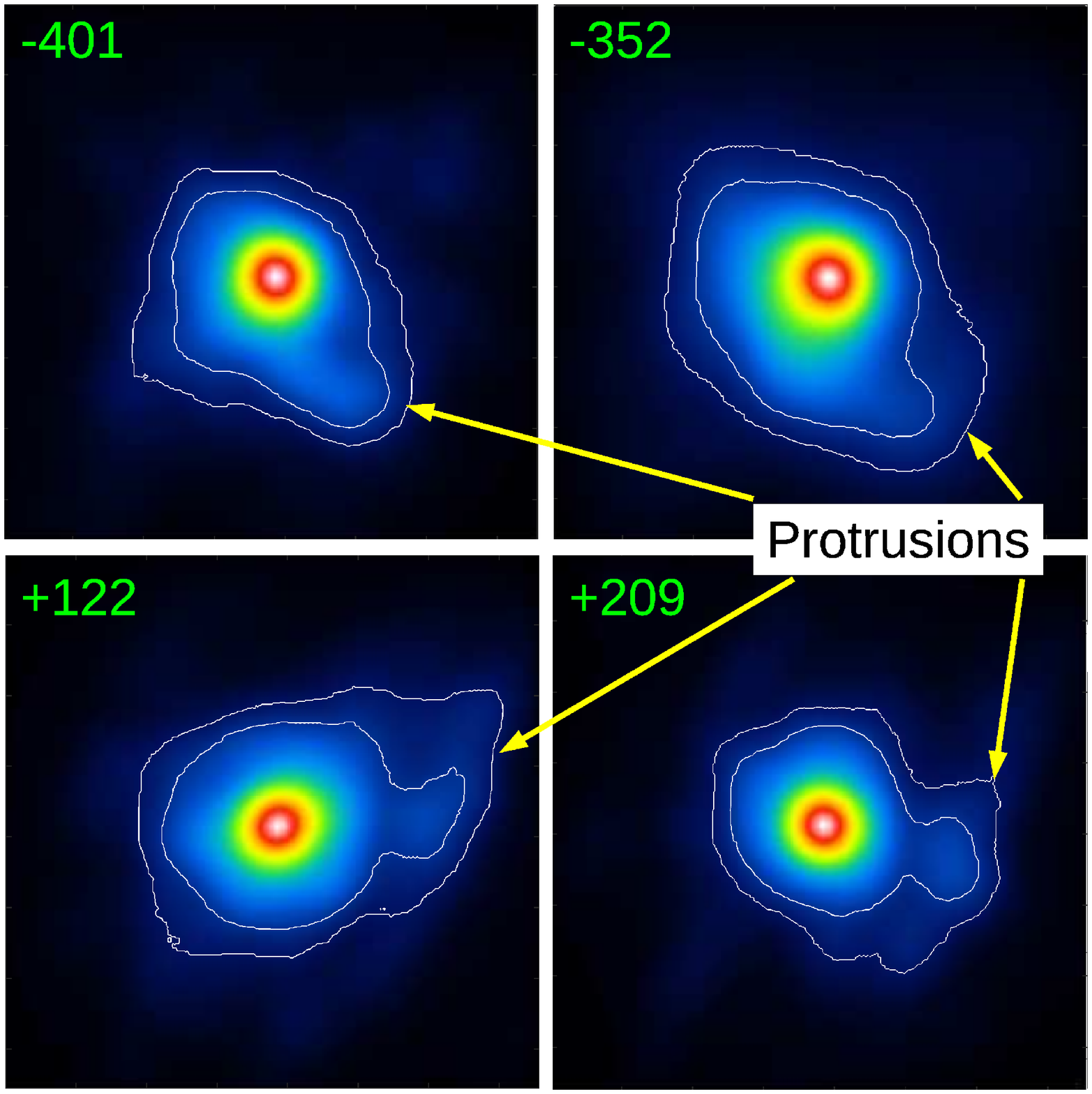}
\caption{Images from \cite{Weigeltetal2016}. The images are oriented with the upper part directed northward and the right side directed eastward. \textit{Upper panel:} (from their figure 5) at a velocity of $-277 \kms$ presents the fan-shaped structure according to their interpretation.
The blue arrow indicates the motion direction, and the two red dots represent the primary and the secondary stars according to their model (which we dispute) at the time of their observations. The bar at the bottom right marks the length of the major axis of the orbit, $30.9 \au$ from \cite{Maduraetal2012}.
The dashed-white ellipse marks the orbital orientation that we propose, where the secondary star is closest to us at periastron passage. 
Note that the orbit shapes have not been corrected for the inclination angel $i\simeq41^{\circ}$.
\textit{Lower panel:} the images at different velocities, given in the insets in units of $\kms$, taken from their figure A.1. Note the different protrusions in the different directions, suggesting that the fan-shaped structure is not unique. 
We add insets with arrows pointing at specific features we refer to in the text. 
The side length of all images is $118 \au$.
For each of the four sub-panels we add contours to emphasize the protrusions.
} 
\label{fig:WeigeltImages}
\end{figure}

We find no justification for their identification of the fan-shaped structure with a colliding winds fan. In Fig. \ref{fig:WeigeltImages} we present images taken from \cite{Weigeltetal2016}. The upper panel presents what they interpret as the fan-shaped structure as appears in the $-277 \km \s^{-1}$ velocity image. In the lower panels we present images at other velocities. The images at these four velocities, as well as other velocities presented in figures A.1 and B.1 of \cite{Weigeltetal2016}, show protrusions that are as large/long as the two legs of the fan-shaped structure identified by those authors, but that are pointed in many other directions.  

Fig. \ref{fig:WeigeltImages} reveals several problems with the claim of \cite{Weigeltetal2016} that the fan-shaped structure results from orientation of the colliding winds of the binary system in $\eta$ Car. 
(1) As already mentioned, there are many features observed in the very same Br~$\gamma$ line sliced radial velocity bins at different line-of-sight velocities, that are of the same spatial size and intensity as the two legs (sides) of the fan-shaped structure. However, these protrusions point in many different directions. Namely, the fan-shaped structure is not unique. 
(2) Even the two legs of the fan-shaped structure itself do not show declining intensity as the distance from the center increases. Instead, the intensity is quite uniform, or even increases to an intensity peak as we mark in the first panel of Fig. \ref{fig:WeigeltImages}. 

We conclude that the fan-shaped structure is not unique, but rather there are scattered blobs along different directions. Two of these blobs form protrusions that \cite{Weigeltetal2016} identify as coming from the colliding winds. We find no justification for that, and hence we dispute their conclusion on the orientation of \etc.

\section{THE DISRUPTED MOLECULAR TORUS}
\label{sec:torus}
\subsection{Properties of the Torus}
\label{subsec:torus:properties}

\cite{Smithetal2018} use the ALMA and detected molecular condensations which they claim to constitute a CO torus around $\eta$ Car. They find that this CO torus misses mass on the side closer to us (the near side or blue shifted side). They then write that the missing material side matches the apastron direction of the secondary star (according to the model we dispute here), and from that conclude that the secondary star disrupted the mass-missing portion of the torus after the latter had been ejected.
Note that the mass is missing from the currently observed torus, but \cite{Smithetal2018} propose that the missing mass did exist in the original torus, and was removed by the secondary star during the first orbit counted from the ejection of the torus. They estimate the total mass in the torus to be $\approx 0.2$--$1 \rmModot$.
We later raised the possibility that the missing-mass side coincides with the periastron side of the secondary star, and the missing mass is due to the mechanism of the GE rather than to a later clearing of the mass-missing part of the torus. 
According to \cite{Smithetal2018} the CO torus expands with a velocity of $v_{\rm to} \simeq 123 \km \s^{-1}$, and it is located at a distance of about $4400 \au$ from the binary system (the torus is thick and composed of several clumps, so its exact distance is not well defined). From these they deduce the age at 2017 to be $170 \pm 15 \yr$, indicating ejection during the GE. 

\cite{Smithetal2018} adopt the orientation where at apastron the companion is toward us and suggest that after the ejection of the Homunculus the companion may have disrupted the expanding torus near apastron. The momentum flux in the torus is much larger than that blown in the wind of the secondary star within the interaction time. For an interaction time of, say, half a year, the momentum discharge flux in the entire secondary wind is $0.5 \yr \times \dot M_2 v_2 \simeq 0.015 \rmModot \km \s^{-1}$, while that in the torus is $M_{\rm to} v_{\rm to} \ga 25  \rmModot \km \s^{-1}$. So the wind blown by the secondary star has practically no influence on the torus.
 
The secondary can accrete mass from the torus. Its radius of gravitational influence, i.e., the Bondi-Hoyle-Lyttleton (BHL) or accretion radius, is about $R_{\rm acc} = 2GM_2/v^2 \approx 5.3 (M_2/30 \rmModot) \au$, where $v\simeq 100 \km \s^{-1}$ is the relative velocity between the secondary and the outflowing gas. This is not negligible even around the apastron distance of $\approx30 \au$. This implies that the secondary star gravitationally deflects the outflowing torus toward its location, and accretes part of this mass.
However, despite the accretion of mass from the torus, there will be no missing-mass torus segment because, as we show below, the deflected mass that is not accreted forms a dense column behind the secondary star. 

Another problem we find with the disrupted-by-secondary explanation for the missing-mass segment in the torus is that the CO torus is not uniform, with missing mass along one segment, but is rather composed of several dense clumps along the red shifted (far) side.
\cite{Artigau2011} have already pointed out that structures that appear coherent in low resolution can be  revealed as composed of clumps when observed at higher resolution. For example, the claimed ``butterfly nebula'' in \etc which was claimed by \cite{Smith2006} to be part of a disrupted equatorial torus. was revealed by \cite{Artigau2011} to be a collection of filaments and clumps.
We therefore consider the torus observed by \cite{Smithetal2018} not to be a torus at all but rather a collection of clumps at different densities.
Nevertheless, the following analysis we perform here shows that whether a torus or a collection of gas condensations had been ejected, it would not have its missing mass on the apastron side but rather on the periastron side.

\subsection{Secondary-Torus Interaction}
\label{subsec:torus:interations}

To study the interaction of the secondary star with the expanding torus shortly after the latter's ejection we performed a three-dimensional hydrodynamical simulation using the \texttt{Flash} v4.5 code \citep{Fryxelletal2000}. We set up a simulation that aims to show that the secondary gravity focuses the gas in the torus rather than creating a hole in it. The uncertainties are quite large regarding the properties of the torus at the time it was ejected (mass, dimensions, clumpiness) so there is no point in keeping all other parameters accurate, or treating high-order effects; nor do we think that this is the process that took place during the GE. We therefore proceed with a proof-of-concept simulation rather than a fully detailed one.

The terminal velocity of the torus is $v_{\rm to} \simeq 123 \km \s^{-1}$. This is much slower than the velocity of the secondary star near periastron, but larger than the velocity at which secondary moves away from the primary at later phases in the orbit. The torus that we assume to have been ejected during a periastron passage  catches up with the secondary star. This occurs $\approx$nine months after periastron passage, when the orbital separation is $\approx 20 \au$. 
At this distance the gravitational influence of the primary star is small, i.e., the orbital speed is smaller than the velocity of the torus. We therefore neglect the gravity of the primary star and include only that of the secondary star.

In our simulation we radially eject a torus of mass $M_{\rm to} = 1 \rmModot$ and with an outward velocity of  $v_{\rm to} = 123 \km \s^{-1}$.
Prior to the ejection of the torus the entire volume is filled with undisturbed primary wind ($\dot{M}_1=6 \times10^{-4} \msyr$, $v_1=500 \kms$) which is of much lower density than the torus.
The initial dimensions of the torus are $R_1<r<3R_1$, where $R_1=180 \rmRodot$ is the radius of the primary, and it fills the solid angle within latitudes $30^\circ$ from both sides of the equator.
As justified, the gravity of the primary is neglected and the secondary gravity is taken as that of a point mass of $M_2=30 \rmModot$, as in the conventional model for $\eta$ Car (see \citealt{KashiSoker2016Massive}).
We neglect the secondary motion and place it still at a distance of $20 \au$ from the primary.
Self-gravity of the torus is negligible and so we do not include it in our simulation.
In our micro-physics treatment we include radiative cooling according to \cite{SutherlandDopita1993}, taking solar metalicity and helium-rich gas with fraction $Y=0.5$.
The secondary environment is treated as a sink, and whatever material reaches the cells inward to $R_{\rm acc} \simeq 5.5 \au$ around the secondary with velocity vector pointed inward is regarded as accreted gas and is removed from the simulation. We do not increase $M_2$ in response to the accreted gas because the increase is insignificant.

\begin{figure*}
\centering
\includegraphics
[trim= 0.1cm 0.5cm 0.2cm 1.1cm,clip=true,width=0.85\textwidth]{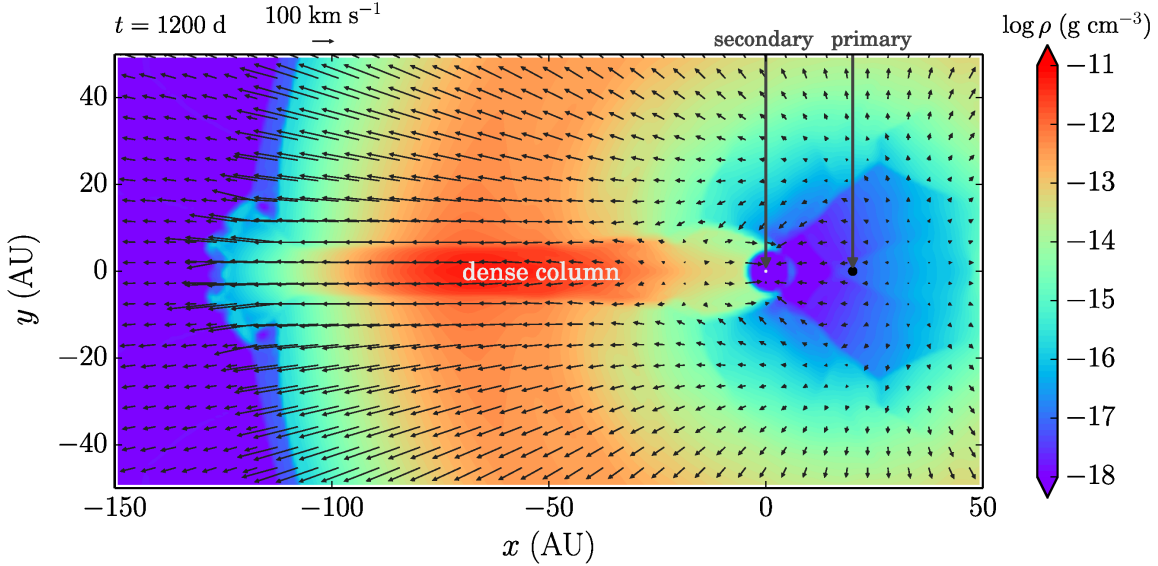} 
\caption{Density map with velocity vectors in the equatorial ($x-y$) plane of our 3D simulation of torus ejection at $t=1200$ days after the ejection of the torus.
The torus is ejected from the primary star in the section of $-30^\circ < \theta < 30^\circ$, where $\theta$ is the latitude (measured from the equatorial plane). The primary star that is located at $(20 \au, 0, 0)$ and the secondary star at $(0, 0, 0)$. 
We label the high-density accretion column and its continuation behind the secondary.
} 
\label{fig:simulation}
\end{figure*}

\begin{figure*}
\centering
\includegraphics
[trim= 0.9cm 2.5cm 2.5cm 1.4cm,clip=true,width=0.85\textwidth]{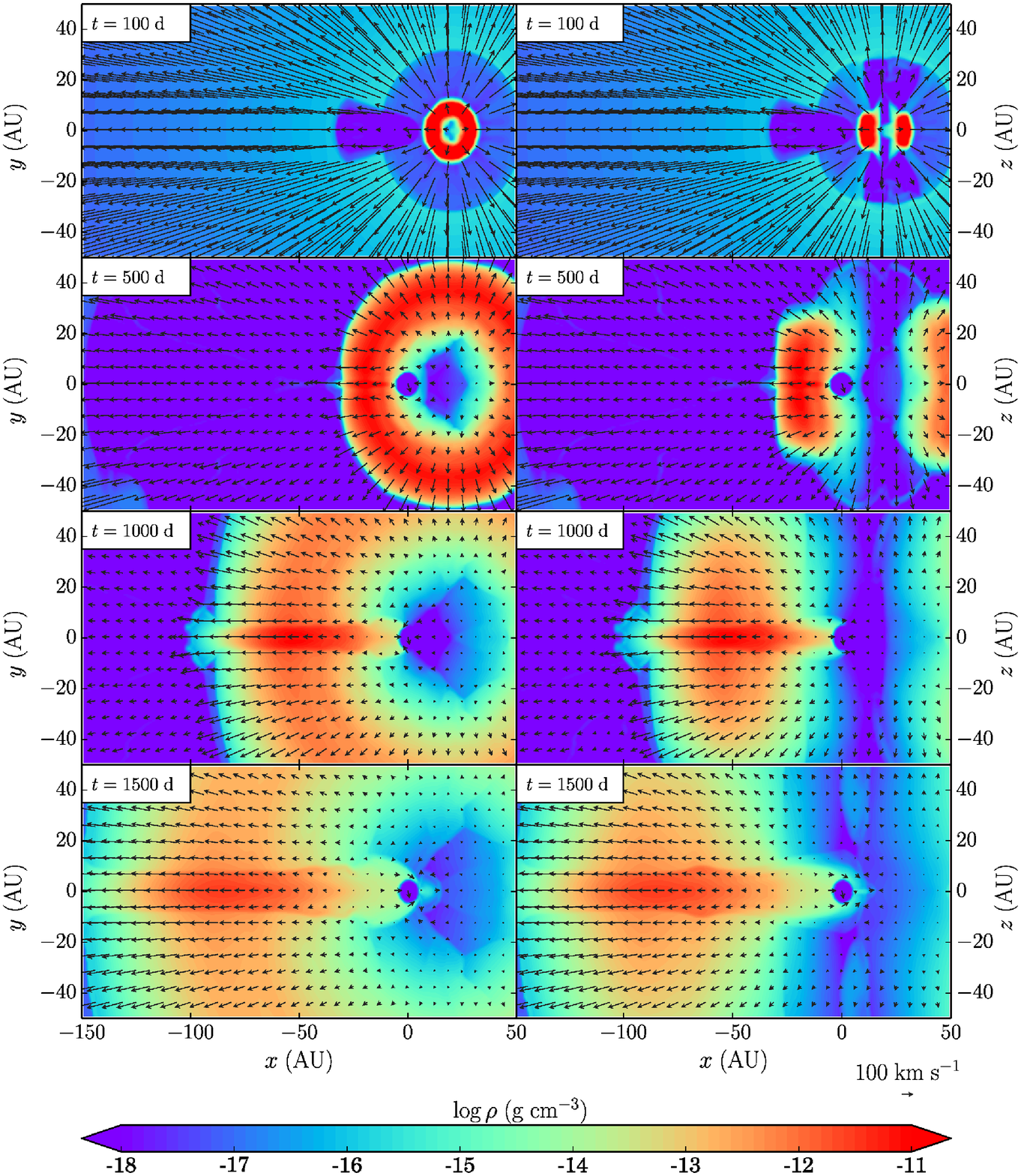} 
\caption{Density maps with velocity vectors at four times.
In the left column we present the flow in the equatorial plane ($x$--$y$ ), and in the right column, in a meridional plane that contains the two stars.    
Despite the accretion of mass onto the secondary star, a dense region is formed behind it. There is no mass-missing region in the torus at later times. } 
\label{fig:simulation_8panel}
\end{figure*}

In Fig. \ref{fig:simulation} we present the density and velocity maps in the equatorial plane at $t=1200$~days. In Fig. \ref{fig:simulation_8panel} we present the density and velocity maps at four times in both the equatorial plane (left column) and a meridional plane that contains the two stars (right column).  
As expected, the outer part of the torus crosses the secondary star in about $260$ days. This time interval is similar to the ballistic time, namely the torus has not slowed down due to the pre-existing primary wind, and launching the torus at the observed velocity was a reasonable simulation setup.
%

Even though the stream lines far from the accreting body of the gas in the expending torus are not parallel to each other as in the classical BHL flow, the properties of the flow that we see in the figures are qualitatively similar to the classical BHL flow that assumes plane-parallel flow \citep{HoyleLyttleton1939, BondiHoyle1944}.
We find that the gravity of the secondary star deflects the gas in the expanding torus and forms of a dense column behind the secondary as the torus passes the secondary. The part of the dense column that is close to the secondary star, within a stagnation point located $\approx R_{\rm acc}$ from the secondary, is accreted onto the secondary star at early times, and the far part of the dense column continues to flow outward.

\begin{figure*}
\centering
\includegraphics
[trim= 0.1cm 0.5cm 1.2cm 1.3cm,clip=true,width=0.85\textwidth]{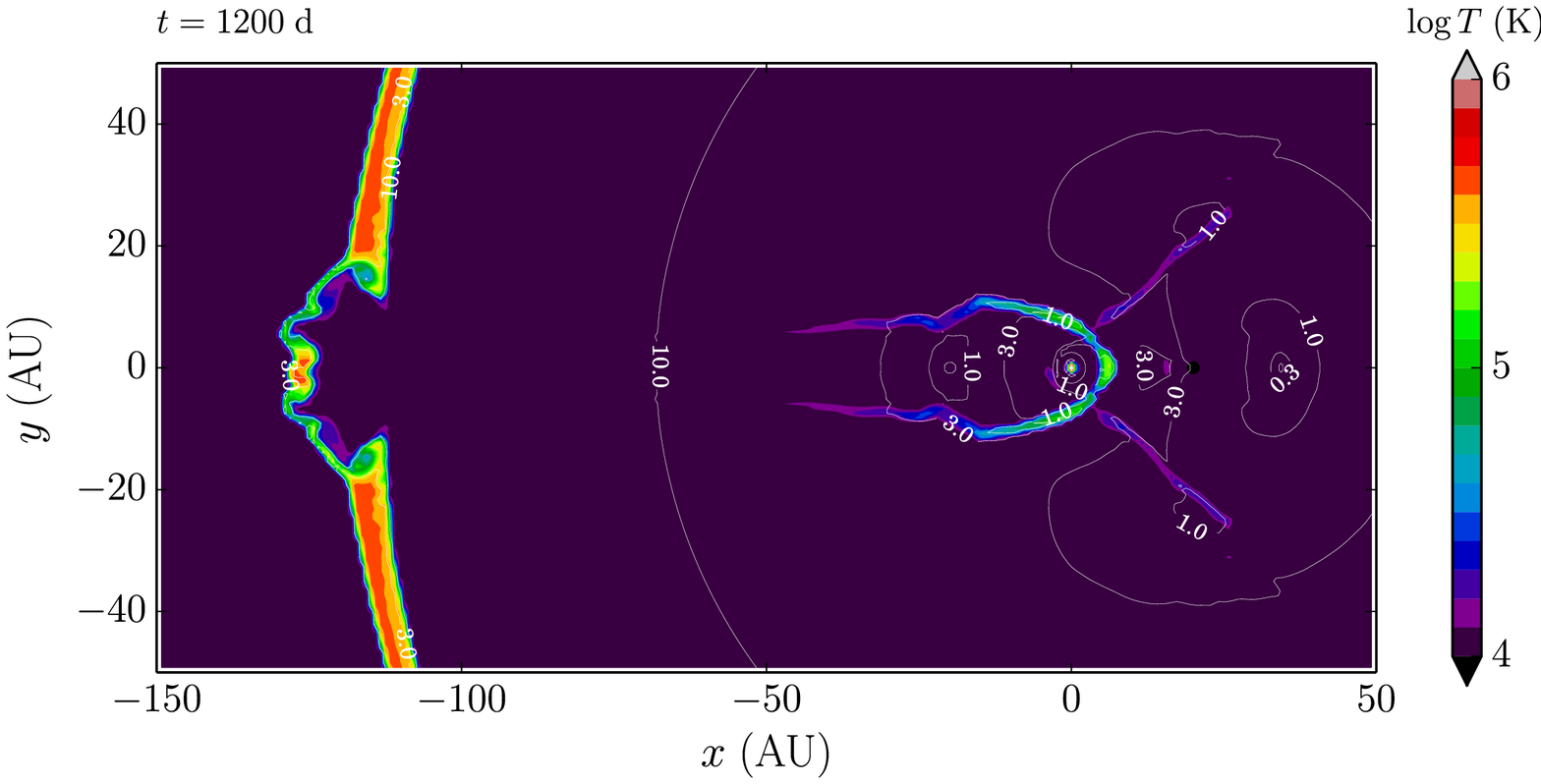} 
\caption{Temperature map with superimposed Mach number contours in the same plane at the same time as in Fig. \ref{fig:simulation}.}
\label{fig:simulation_tempandmach}
\end{figure*}

In Fig. \ref{fig:simulation_tempandmach} we present the temperature map in the equatorial plane at $t=1200$~days. While most of the torus cools to a temperature of $T<10^4 \K$, the edges of the dense column have temperatures in the range of few~$\times 10^4$ -- a few~$\times 10^5 \K$, with hotter temperatures close to the secondary at the accreting part of the column. The accretion shock in front of the secondary is prominent.
%

The inner parts of the torus slow down while the outer parts keep expending at the same velocity, which makes the torus radius grow with time.
This occurs because there is expansion of the entire torus outward plus self-expansion as it is denser than its environment. The combined effect is that the inner radius of the torus propagates outward more slowly than the outer radius; namely, its radial thickness is increasing. 
Therefore, accretion from the closer part of the column and inner parts of the torus continues even after most of the torus has already passed the secondary.
The secondary star have accreted a total mass of $M_{\rm acc} \simeq 0.15 \rmModot$ from the expanding torus that crosses it. 

Had we taken the motion of the secondary into account the dense accretion column formed by the deflected torus would have been somewhat wider and of somewhat lower density, which would not have any significance to our conclusion. It is therefore clear that the secondary star causes a density enhancement behind it rather than a density depression.

\section{DISCUSSION AND SUMMARY}
\label{sec:summary}
 
In this study we critically examined several recent papers that suggest that the orientation of the binary system $\eta$ Car has $\omega \simeq 240^\circ-270^\circ$, namely, it is such that the secondary star is closer to us during most of the orbit, i.e., at apastron, and it is behind the primary star during the short periastron passage. We found problems with arguments for this orientation that are based on absorption lines (section \ref{subsec:abslines}), on the \ion{He}{2}~$\lambda4686$ emission line (section \ref{subsec:HeII4686line}), on the clumps in the vicinity of the binary system (section \ref{sec:fan}), and on the missing-mass segment in the torus that was ejected during the GE (section \ref{sec:torus}). 
     
Our numerical simulations of the evolution of the torus that \cite{Smithetal2018} studied show that the orientation with $\omega \simeq 240^\circ-270^\circ$ would result in a dense torus segment, rather than a missing-mass segment. More generally, if the torus is ejected in a more or less symmetric manner, our results rule out an orientation with $180^\circ \la \omega \la 360^\circ$. We therefore return to discuss the orientation with $\omega \simeq 90^\circ$, i.e., the secondary star is away from us during most of the orbit, and it is in front of the primary star during periastron passage (up to the inclination of the orbit).

In previous papers we have already presented arguments to show that absorption and emission lines 
(\citealt{KashiSoker2007,KashiSoker2008,KashiSoker2016Massive}), the hydrogen column density toward the binary system as deduced from X-ray absorption (\citealt{KashiSoker2009b,KashiSoker2009d}), emission from blobs around the binary system \citep{SokerKashi2012}, and other observations, all support the orientation of $\omega \simeq 90^\circ$, i.e., secondary away from us during most of the orbit and in front of the primary at periastron.

We now discuss the possible implications of the missing-mass segment in the torus as reported by \cite{Smithetal2018}. 
The relevant properties of the torus and their implications are as follows. 
(1) It is clumpy. This hints at a violent and/or unstable process. We attribute it to a mass ejection in a very short time associated with one or two periastron passages. 
(2) There is a mass concentration on the far side of the torus \citep{Smithetal2018}. As we showed in section \ref{subsec:torus:interations}, when the torus catches up with the secondary star, the outcome is a dense column behind it. We attribute the concentration of mass to such an interaction. Note that, due to the motion of the secondary star and clumpy mass ejection in the torus, the column will not be as smooth and narrow as in our simulation that we present in Figures \ref{fig:simulation} and \ref{fig:simulation_8panel}.
(3) There is a mass-missing segment on the side of the torus closer to us. 

\begin{figure}
\centering
\includegraphics[trim= 1.7cm 1.1cm 1.7cm 0.1cm,clip=true,width=0.99\columnwidth]{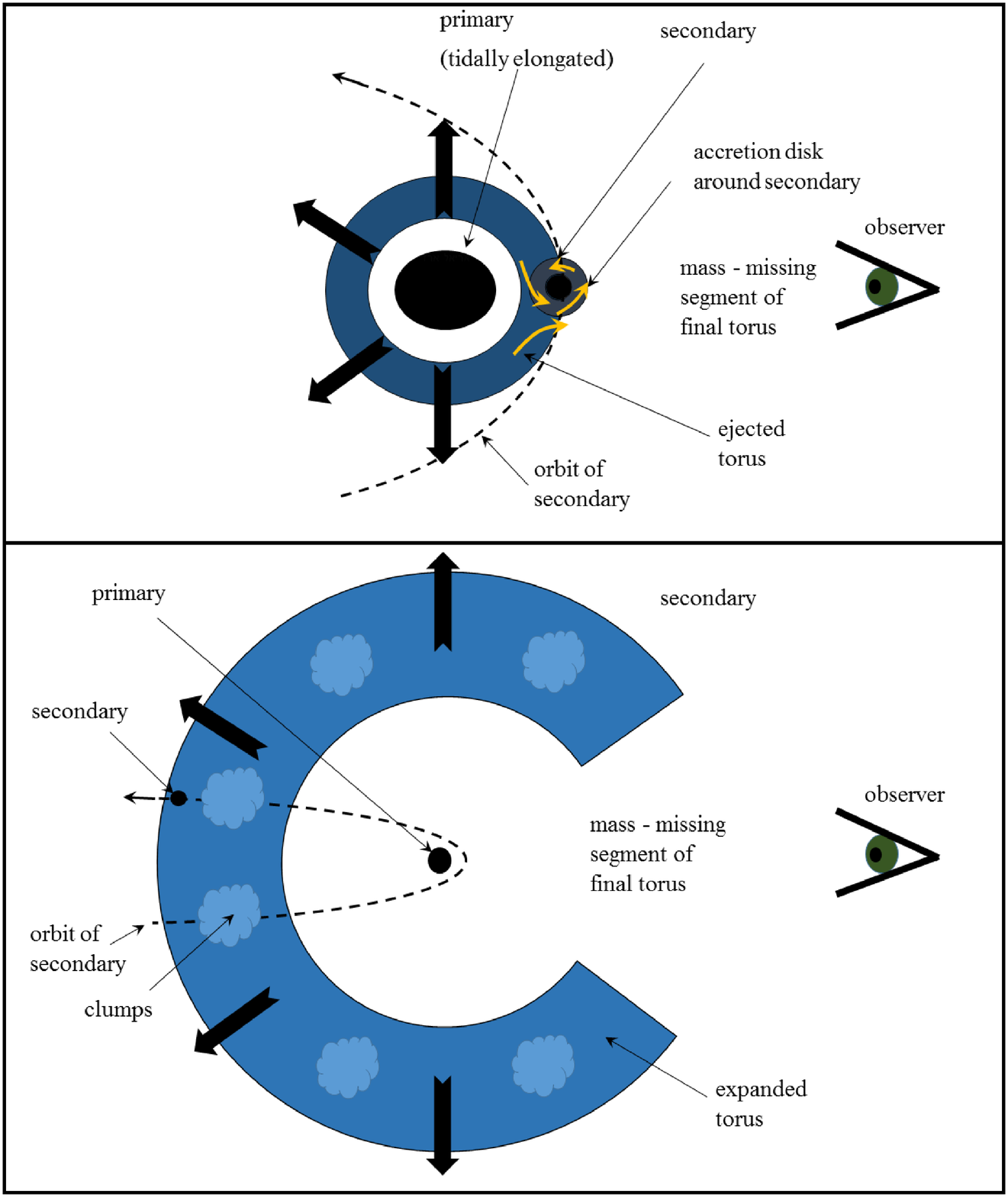} 
\caption{Schematic drawing (not to scale) of the formation of the mass-missing segment in the ALMA-molecular torus or collection of condensations.
\textit{Upper panel:} the torus (or a collection of clumps) is being launched while the secondary is close to periastron passage. The secondary accretes mass from the torus.
Accretion onto the secondary in the acceleration zone of the torus removes mass from the torus and forms an accretion disk. Since accretion takes place in the acceleration zone of the torus and in a time scale shorter than its outflow time, no dense column is formed behind the secondary star. 
\textit{Lower Panel:} the torus expands. The accretion of mass onto the secondary star that occurred at periastron prevents mass from flowing out from the system in the general periastron direction, as marked. Later, the torus on the apastron side catches up with the secondary star, which accretes some more mass and deflects the torus to reduce its homogeneity, i.e., forming a dense part in the apastron direction (section \ref{subsec:torus:interations}). }
\label{fig:schematic}
\end{figure}

We attribute the mass-missing segment of the torus or gas clumps to the presence of the secondary on that direction during the periastron passage in the GE. Firstly, the mass was ejected, most likely, to the other side of the primary, where there is a tidal bulge, namely on the other side of the secondary near periastron passage.
Second, the secondary star accreted a large amount of mass during its periastron passage (\citealt{KashiSoker2010a}), hence preventing mass ejection in its direction relative to the primary star at that time.
Accretion onto the secondary took place in the acceleration zone of the torus. Instead of BHL-like accretion, it better resembled Roche-Lobe overflow (RLOF) accretion (see the discussion on the different modes of accretion in $\eta$ Car in \citealt{KashiSoker2009c}).
The accretion prevents the construction of a full torus, removes mass from it and forms an accretion disk.
We schematically present our proposed scenario in Fig. \ref{fig:schematic}.
%

With our conclusion of the last point we expand the question of the orientation of \etc to one with far-reaching implications. It is not just a question of studying the specific case of \etc and the origin of different lines and absorbing gas. The orientation of \etc is related to the accretion processes during the periastron passage of the GE. 
The luminous peaks of the GE were what are now termed supernova impostors \citep{HumphreysMartin2012}, namely, outbursts with a timescales and a luminosities close to those of supernovae. These impostors form one group in the heterogeneous group of ILOTs \citep{KashiSoker2016}. We attribute the powering of ILOTs to high accretion rate onto a main sequence (MS) or a slightly evolved off the MS star, which we term the high-accretion-powered ILOTs (HAPI) model \citep{KashiSoker2016}.
The luminosity peaks of the GE, according to the HAPI model, were powered by the accretion of large amount of mass on to the secondary star during periastron passage in the GE. 

Our suggestion that the mass-missing segment of the torus results from the accretion process on to the secondary star during the GE periastron passage strengthens our earlier claim that the secondary star accreted lots of mass and that this was the largest energy source of the GE \citep{KashiSoker2010a}. This in turn strengthens the more general model according to which many ILOTs are powered by accretion onto a secondary star \citep{KashiSoker2010b}.

\vspace{0.3cm}
We thank Gerd Weigelt for helpful discussions on the fan-shaped structure.
We greatly appreciate comments from an anonymous referee that helped to improve the paper.
This research was supported by the Asher Fund for Space Research at the Technion, and the Israel Science Foundation.
A.K. acknowledges support from the R\&D authority in Ariel University and the Rector of Ariel University.
This work used the Extreme Science and Engineering Discovery Environment (XSEDE), which is supported by National Science Foundation grant number ACI-1548562.
This work used the Extreme Science and Engineering Discovery Environment (XSEDE) TACC/Stampede2 at the service-provider through allocation TG-AST150018.
This work was supported by the Cy-Tera Project, which is co-funded by the European Regional Development Fund and the Republic of Cyprus through the Research Promotion Foundation.

\end{document}